\documentclass[letterpaper,twocolumn,pra,preprintnumbers,showpacs]{revtex4-1}
\usepackage{graphicx,amsmath, amssymb, bm,epstopdf, mathtools }

\begin{document}
\title{Transfer of temporal coherence in parametric down-conversion}
\author{Girish Kulkarni, Prashant Kumar$^\dagger$, and Anand K. Jha}
\email{akjha9@gmail.com;$^\dagger$Currently at Stanford University, USA}
\affiliation{Department of Physics, Indian Institute of Technology, Kanpur 
208016, India}

\begin{abstract}
We show that in parametric down-conversion the coherence
properties of a temporally partially coherent pump field get
entirely transferred to the down-converted entangled two-photon
field. Under the assumption that the frequency-bandwidth of the
down-converted signal-idler photons is much larger than that of
the pump, we derive the temporal coherence functions for the
down-converted field, for both infinitely-fast and time-averaged
detection schemes. We show that in each scheme the coherence
function factorizes into two separate coherence functions with one
of them carrying the entire statistical information of the pump
field. In situations in which the pump is a Gaussian Schell-model
field, we derive explicit expressions for the coherence functions.
Finally, we show that the concurrence of time-energy-entangled
two-qubit states is bounded by the degree of temporal coherence of
the pump field. This study can have important implications for
understanding how correlations of the pump field manifest as
two-particle entanglement as well as for harnessing energy-time
entanglement for long-distance quantum communication protocols.
\end{abstract}

\maketitle

\section{Introduction}

Coherence and entanglement are intimately related concepts. The
recent attempts at developing a resource-based theory of coherence
also reveal such relations \cite{vogel2014pra, baumgratz2014prl,
girolami2014prl, chitambar2016prl}. One of the physical processes
in which the relations between coherence and entanglement can be
systematically explored is parametric down-conversion (PDC)---a
nonlinear optical process in which a pump photon interacts with a
nonlinear crystal to produce a pair of entangled photons, termed
as signal and idler \cite{burnham1970prl}. Using the PDC photons,
coherence and entanglement effects have been observed in several
degrees of freedom including polarization \cite{brendel1995pra},
time-energy \cite{hong1987prl, zou1991prl, herzog1994prl,
pittman1996prl, jha2008pra, franson1989prl, brendel1991prl,
jha2008prl, brendel1999prl, thew2002pra}, position and momentum
\cite{fonseca1999pra, neves2007pra, jha2010pra}, and orbital
angular momentum (OAM) \cite{nagali2009natphot, jha2010prl,
pires2010prl, jha2011pra}.

There have been several studies on how coherence and entanglement
properties of the down-converted field are affected by different
PDC setting and pump field parameters \cite{hong1985pra,
rubin1996pra, monken1998pra, joobeur1996pra, ribeiro1997pra,
fonseca1999pra, saleh2005prl}. However, regarding how the
intrinsic correlations of the pump field get transferred to
manifest as two-photon coherence and entanglement, there have been
efforts mostly in the polarization and spatial degrees of freedom
\cite{monken1998pra, walborn2004pra, jha2010pra, kulkarni2016pra}.
In the spatial degree of freedom, a very general spatially
partially coherent field was considered and it was shown that the
spatial coherence properties of the pump field get entirely
transferred to that of the down-converted two-photon field
\cite{jha2010pra}. However, in the temporal degree of freedom, the
effects due to the temporal correlations of the pump field have
only been studied in two limiting situations: one, in which the
constituent frequency components are completely correlated
(fully-coherent pulsed field) \cite{grice1997pra, keller1997pra,
brendel1999prl,tittel2000prl, mikhailova2008pra, inagaki2013optexp} and the other, in
which the constituent frequency components are completely
uncorrelated (continuous-wave field) \cite{hong1987prl,
zou1991prl, herzog1994prl, pittman1996prl, jha2008pra,
franson1989prl, brendel1991prl, jha2008prl, ou1989pra,
rubin1994pra, milonni1996pra, kwon2009optexp, nasr2008prl, okano2015scirep, tanaka2012optexp}. In
this article, we study the coherence transfer in PDC for a general
temporally partially coherent pump field and explicitly quantify
this correlation transfer for the special case of a partially
coherent Gaussian Schell-model field \cite{paakkonen2002optcomm},
in which the correlations between the constituent frequency
components have a Gaussian distribution.

The paper is organized as follows. In Sec.~II, we consider a
general temporally partially coherent pump field and show that its
temporal coherence properties get entirely transferred to the
down-converted two-photon field. We work out the two-photon
temporal coherence functions for both infinitely-fast and
time-averaged detection schemes and show that in each scheme the
coherence function factorizes into two separate coherence
functions with one of them carrying the entire statistical
information of the pump field. In Sec.~III, we show that the
entanglement of time-energy entangled two-qubit states is bounded
by the degree of temporal coherence of the partially coherent pump
field. We present our conclusions in Sec.~IV.

\section{Tranfer of Temporal coherence  in PDC}

\subsection{Detection with infinitely fast detectors}

\begin{figure}[t!]
\includegraphics[width=88mm,keepaspectratio]{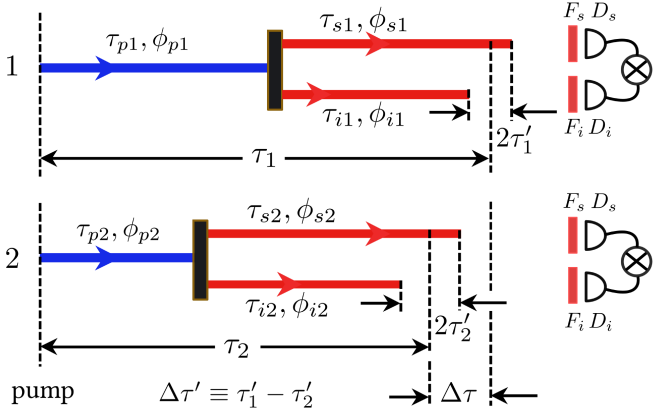}
\caption{(color online). Schematic representation of two-photon
interference using the two-photon path diagrams. Alternatives 1 and 2 are the two pathways by which a pump photon is down-converted and the down-converted photons are detected at single-photons detectors $D_{s}$ and $D_{i}$.
}\label{two-photon-path-diagram}
\end{figure}

We follow the formalism worked out in Ref.~\cite{jha2008pra} and
represent a general two-alternative two-photon interference of the
PDC photons by the two-photon path diagrams shown in
Fig.~\ref{two-photon-path-diagram}. The pump is a general
temporally partially coherent field. Alternatives 1 and 2 are the
two pathways by which a pump photon is down-converted and the
down-converted signal and idler photons are detected in
coincidence at single-photon detectors $D_{s}$ and $D_{i}$,
respectively. There are six independent time parameters in this
setting. The subscripts $p$, $s$ and $i$ denote the pump, signal
and idler respectively. We adopt the convention that a signal
photon is the one that arrives at detector $D_{s}$ while an idler
photon is the one that arrives at detector $D_{i}$. The symbol
$\tau$ denotes the traversal time of a photon while $\phi$ denotes
the phase, other than the dynamical phase, accumulated by a
photon. Thus, $\tau_{s1}$ denotes the traversal time of the signal
photon in alternative $1$, etc. The various signal, idler and pump
quantities are used to define the following parameters:
\begin{align}
\Delta\tau  \equiv  \tau_{1}-\tau_{2} &\equiv
\left(\tau_{p1}+\frac{\tau_{s1}+\tau_{i1}}{2}\right)-\left(\tau_{p2}+\frac{\tau_
{s2}+\tau_{i2}}{2}\right),\notag \\ \notag \Delta\tau' \equiv
\tau'_{1}-\tau'_{2} &\equiv
\left(\frac{\tau_{s1}-\tau_{i1}}{2}\right)-\left(\frac{\tau_{s2}-\tau_{i2}}{2}\right),\\
\Delta\phi \equiv \phi_1-\phi_2&\equiv
\left(\phi_{p1}+\phi_{s1}+\phi_{i1}\right)-\left(\phi_{p2}+\phi_{s2}+\phi_{i2}
\right).\label{definitions}
\end{align}
The parameters defined above are identical to those defined in
Ref.~\cite{jha2008pra}, except for $\tau'_{1},\tau'_{2}$ and
$\Delta\tau'$, which have been scaled down by a factor of 2. It is
found that this rescaling imparts the equations in this paper a
neat and symmetric form.
The two-photon state $|\psi\rangle_1$ produced in alternative $1$
in the weak-downconversion limit  is given by \cite{ou1989pra,
wang1991pra, grice1997pra}:
\begin{multline}\label{pdc_state}
|\psi\rangle_{1}= A_{1}\iint d\omega_{s} d\omega_{i}V(\omega_{p})
\Phi_{1}(\omega_{s},\omega_{i})\\\times
e^{i(\omega_{p}\tau_{p1}+\phi_{p1})}|\omega_s\rangle_{\omega_{s1}}|\omega_i
\rangle_{\omega_{i1}},
\end{multline}
where $V(\omega_{p})$ is the random, spectral amplitude of the
pump field at frequency $\omega_{p}$ and
$\Phi_{1}(\omega_{s},\omega_{i})$ is the phase-matching function
in alternative 1. The two-photon state $|\psi\rangle_2$ in
alternative $2$ can be similarly defined. The complete two-photon
state $|\psi\rangle$ at the detectors is the sum of the two-photon
states in alternatives $1$ and $2$ and can be written as
$|\psi\rangle=|\psi\rangle_{1}+|\psi\rangle_{2}$. The
corresponding density matrix $\rho$ of the state at the detectors
is therefore:
\begin{equation}\label{rho}
 \hat{\rho}=\big\langle |\psi\rangle\langle\psi| \big\rangle.
\end{equation}
Here $\langle\cdots\rangle$ represents an ensemble average over
infinitely many realizations of the two-photon state.

We now denote the positive frequency parts of the electric fields
at detectors $D_{s}$ and $D_{i}$ by $\hat{E}^{(+)}_{s}(t)$ and
$\hat{E}^{(+)}_{i}(t)$, respectively, and write them as
\begin{subequations}\label{field_operators}
\begin{align}
&\hat{E}_{s}^{(+)}(t)=\kappa_{s1}\hat{E}_{s1}^{(+)}(t-\tau_{s1})+\kappa_{s2}\hat
{E}_{s2}^{(+)}(t-\tau_{s2}),\\
&\hat{E}_{i}^{(+)}(t)=\kappa_{i1}\hat{E}_{i1}^{(+)}(t-\tau_{i1})+\kappa_{i2}\hat
{E}_{i2}^{(+)}(t-\tau_{i2}),
\end{align}
\end{subequations}
where $\kappa_{s1(2)}$ and $\kappa_{i1(2)}$ are scalar amplitudes
and where
\begin{align}\label{fields}
\hat{E}_{s1}^{(+)}(t-\tau_{s1})=  &  \notag \\
e^{i\phi_{s1}} \int^{\infty}_{0}
 & d\omega
f_{s1}(\omega-\omega_{s0}) \hat{a}_{s1} (\omega)
e^{-i\omega(t-\tau_{s1})},
\end{align}
is the positive frequency parts of the electric field at detector
$D_s$ in alternative $1$, etc. The function
$f_{s1}(\omega-\omega_{s0})$ is the amplitude transmission
function of the filter $F_{s}$ placed at detector $D_s$, etc. The
filters $F_{s}$ and $F_{i}$ are centered at frequencies
$\omega_{s0}$ and $\omega_{i0}$, respectively, and we assume the
phase-matching condition $\omega_{p0}=\omega_{s0}+\omega_{i0}$,
where $\omega_{p0}$ is the central frequency of the pump field
$V(\omega_p)$. The coincidence count rate
$R_{si}^{(2)}(t_{s},t_{i})$ of the two detectors is the
probability per (unit time)$^{2}$ that a signal photon is detected
at time $t_{s}$ and the corresponding idler photon is detected at
time $t_{i}$, and it is given by
$R^{(2)}_{si}(t_{s},t_{i})=\mathrm{Tr}\{\hat{\rho}\,\hat{E}^{(-)}_{s}(t_{s})\,\hat{E}^{
(-)}_{i}(t_{i})\hat{E}^{(+)}_{i}(t_{i})\,\hat{E}^{(+)}_{s}(t_{s})\}$
\cite{glauber1963pr}. Using the definitions and expressions of
Eqs.~(\ref{definitions})-(\ref{fields}), we evaluate
$R^{(2)}_{si}(t_{s},t_{i})$ to be
\begin{subequations}\label{interference equation}
\begin{multline}
R^{(2)}_{si}(t_{s},t_{i})=  \\ \kappa_1^2 R^{(2)}(t_{s},t_{i},
\tau_{s1}, \tau_{i1}) +\kappa_2^2 R^{(2)}(t_{s},t_{i}, \tau_{s2},
\tau_{i2})\\+ \kappa_1\kappa_2
 {\Gamma}^{(2)}(t_{s},t_{i}, \tau_{s1},
\tau_{i1}, \tau_{s2}, \tau_{i2}) e^{-i\Delta\phi} + {\rm c.c.},
\end{multline}
where $\kappa_1=\kappa_{s1}\kappa_{i1}$,
$\kappa_2=\kappa_{s2}\kappa_{i2}$,
\begin{multline}\label{r12}
{\Gamma}^{(2)}(t_{s},t_{i}, \tau_{s1}, \tau_{i1}, \tau_{s2},
\tau_{i2}) =
\mathrm{Tr}\big\{\hat{\rho}\,\hat{E}_{s1}^{(-)}(t_{s}-\tau_{s1})\\
\times\hat{E}_{i1}^{ (-)}(t_{i} -\tau_{i1}) \hat{E}_{i2}^{
(+)}(t_{i}-\tau_{i2})\hat{E}_{s2}^{(+)}(t_{s}-\tau_{s2})\big\},
\end{multline}
{\rm and}
\begin{multline}
R^{(2)}(t_{s},t_{i}, \tau_{s1}, \tau_{i1})=
{\Gamma}^{(2)}(t_{s},t_{i}, \tau_{s1}, \tau_{i1}, \tau_{s1},
\tau_{i1}).
\end{multline}
\end{subequations}
Eq.~(\ref{interference equation}) is the interference law for the
two-photon field. The first and the second terms are the
coincidence count rates in alternatives 1 and 2, respectively. The
interference term ${\Gamma}^{(2)}(t_{s},t_{i}, \tau_{s1},
\tau_{i1}, \tau_{s2}, \tau_{i2})$ appears when both the
alternatives are present, and it will be referred to as the
two-photon cross-correlation function of the down-converted field.
We now make the assumption that the spectral width
$\Delta\omega_{p0}$ of the pump field is much smaller than the
central frequency $\omega_{p0}$ and the spectral widths of the
phase-matching functions and filter functions. As a result, the
phase-matching and filter functions can be taken to be
approximately constant in the frequency range
$\left(\omega_{p0}-\Delta\omega_{p0}/2,
\omega_{p0}+\Delta\omega_{p0}/2\right)$. This assumption remains
valid for most PDC experiments employing continuous wave pump
field \cite{hong1987prl, zou1991prl, herzog1994prl,
pittman1996prl, jha2008pra, franson1989prl, brendel1991prl,
jha2008prl, kwon2013optexp} and pulsed pump field
\cite{grice1997pra, keller1997pra, brendel1999prl,tittel2000prl,
inagaki2013optexp} and may only be invalid for experiments
employing ultrashort pulsed pump fields \cite{marcikic2002pra,
marcikic2003nature,marcikic2004prl}. We use the relations
$\omega_{p}=\omega_{s}+\omega_{i}$,
$\omega_{d}=\omega_{s}-\omega_{i}$,
$\omega_{p0}=\omega_{s0}+\omega_{i0}$,
$\omega_{d0}=\omega_{s0}-\omega_{i0}$ and define the integration
variables $\bar{\omega}_{p}=\omega_{p}-\omega_{p0}$ and
$\bar{\omega}_{d}=\omega_{d}-\omega_{d0}$. Using
Eqs.~(\ref{definitions})-(\ref{interference equation}), we obtain
after a long but straightforward calculation:
\begin{multline}\label{r12-decomp}
{\Gamma}^{(2)}(t_{s},t_{i}, \tau_{s1}, \tau_{i1}, \tau_{s2},
\tau_{i2}) = \\ \times
{\Gamma}_{p}\left(\tau_1-\frac{t_s+t_i}{2},
\tau_2-\frac{t_s+t_i}{2}\right)\\
\times{\Gamma}_{d}\left(\tau'_1-\frac{t_s-t_i}{2},
\tau'_2-\frac{t_s-t_i}{2}\right),
\end{multline}
where,
\begin{multline}\label{Gamma-p}
{\Gamma}_{p}\left(\tau_1-\frac{t_s+t_i}{2},
 \tau_2-\frac{t_s+t_i}{2}\right)
\\=e^{-i\omega_{p0}\Delta\tau}\iint d\bar{\omega}'_{p} d\bar{\omega}''_{p}
\left\langle\,V^*(\bar{\omega}'_{p}
+\omega_{p0})V(\bar{\omega}''_{p} +\omega_{p0})\right\rangle\\
\times \exp\left[ -i\bar{\omega}'_{p}\left(\tau_{1}-\frac{t_s +
t_i}{2}\right)\right] \exp\left[
i\bar{\omega}''_{p}\left(\tau_{2}-\frac{t_s +
t_i}{2}\right)\right],
\end{multline}
and,
\begin{multline}\label{Gamma-d}
{\Gamma}_{d}\left(\tau'_1-\frac{t_s-t_i}{2},
\tau'_2-\frac{t_s-t_i}{2}\right)
\\ =e^{-i\omega_{d0}\Delta\tau'}\iint d\bar{\omega}'_{d} d\bar{\omega}''_{d}
\left\langle
g_{1}^*(\bar{\omega}'_{d})\,g_{2}(\bar{\omega}''_{d})\right\rangle\\
\times \exp\left[ -i\bar{\omega}'_{d}\left(\tau'_{1}-\frac{t_s -
t_i}{2}\right)\right] \exp\left[
i\bar{\omega}''_{d}\left(\tau'_{2}-\frac{t_s -
t_i}{2}\right)\right],
\end{multline}
with
\begin{align}\notag
g_{1}(\omega)=\Phi_{1}\left(\omega_{s0}+\frac{\omega}{2},\omega_{i0}
-\frac{\omega}{2}\right)
f_{s1}\left(\frac{\omega}{2}\right)\,f_{i1}\left(-\frac{\omega}{2}\right),
\end{align}
etc. The ensemble average
$\left\langle\,V^*(\bar{\omega}'_{p}+\omega_{p0})V(\bar{\omega}''_{p}
+\omega_{p0})\,\right\rangle$ is the cross-spectral density
function of the pump field. It is at once clear from
Eq.~(\ref{Gamma-p}) that the coherence function
${\Gamma}_{p}\left(\tau_1-\frac{t_s+t_i}{2},
\tau_2-\frac{t_s+t_i}{2}\right)$ and the cross-spectral density
function
$\left\langle\,V^*(\bar{\omega}'_{p}+\omega_{p0})V(\bar{\omega}''_{p}
+\omega_{p0})\,\right\rangle$ are connected through the
generalized Wiener-Khintchine relation \cite{mandel&wolf} with
parameters $\tau_{1}-(t_s+t_i)/2$ and $\tau_{2}-(t_s+t_i)/2$. So,
in terms of the two-photon time parameters $\tau_{1}-(t_s+t_i)/2$
and $\tau_{2}-(t_s+t_i)/2$, the coherence function
${\Gamma}_{p}\left(\tau_1-\frac{t_s+t_i}{2},
\tau_2-\frac{t_s+t_i}{2}\right)$ has the same functional form as
that of the cross-correlation function of the pump field. The
function $ \left\langle
g_{1}^*(\bar{\omega}'_{d})\,g_{2}(\bar{\omega}''_{d})\right\rangle$
is also in the form of a cross-spectral density function, and as
is clear from Eq.~(\ref{Gamma-d}), it forms a generalized
Wiener-Khintchine relation with the coherence function
${\Gamma}_{d}\left(\tau'_1-\frac{t_s-t_i}{2},
\tau'_2-\frac{t_s-t_i}{2}\right)$. Therefore, the function
${\Gamma}_{d}\left(\tau'_1-\frac{t_s-t_i}{2},
\tau'_2-\frac{t_s-t_i}{2}\right)$ not only carries all the
information about the phase-matching conditions and the crystal
parameters but also carries information about any statistical
randomness that the down-converted photons go through
\cite{jha2010pra2}. It is interesting to note that any statistical
randomness encountered by the photons after the down-conversion
affects only ${\Gamma}_{d}\left(\tau'_1-\frac{t_s-t_i}{2},
\tau'_2-\frac{t_s-t_i}{2}\right)$ and has no effect on
${\Gamma}_{p}\left(\tau_1-\frac{t_s+t_i}{2},
\tau_2-\frac{t_s+t_i}{2}\right)$. This fact can potentially be used for encoding information in the pump's coherence function and decoding if from the down-converted photons even after the down-converted photons have passed through turbulent media.

We thus find that the two-photon cross-correlation function
factorizes into two separate coherence functions. The coherence
function ${\Gamma}_{p}\left(\tau_1-\frac{t_s+t_i}{2},
\tau_2-\frac{t_s+t_i}{2}\right)$ carries the entire statistical
information of the pump field, and in this way the temporal
correlation properties of the pump photon get entirely transferred
to the down-converted photons. This result is the temporal analog
of the effect described in Ref.~\cite{jha2010pra} in which it was
shown that in PDC the spatial coherence properties of the pump
field gets entirely transferred to the down-converted two-photon
field. However, the present paper extends beyond just establishing
this analogy. For example, in Ref.~\cite{jha2010pra}, the effect
due to the phase-matching function was completely ignored, but in
the present paper, we have included it through the coherence
function ${\Gamma}_{d}\left(\tau'_1-\frac{t_s-t_i}{2},
\tau'_2-\frac{t_s-t_i}{2}\right)$. Moreover, like most
spatial-interference schemes, Ref.~\cite{jha2010pra} does not
employ a detection scheme that involves space-averaging. However,
most time-domain experiments employ time-averaged detection
schemes. Therefore, in the present paper, we also work out how
time-averaged detection schemes affect the temporal coherence
transfer in PDC.

\subsection{Time-averaged detection scheme}

In most experiments, one does not measure the instantaneous
coincidence rate $R_{si}^{(2)}(t_{s},t_{i})$ of
Eq.~(\ref{interference equation}). Instead, one measures the
time-averaged coincidence count rate, averaged over the photon
collection time $T_{\rm pc}$ and the coincidence time-window
$T_{\rm ci}$. The time-averaged two-photon cross-correlation
function ${\bar\Gamma}^{(2)}$ can be found by first expressing
it as
\begin{align}
{\bar\Gamma}^{(2)}&=\left\langle\left\langle
 {\Gamma}^{(2)}(t_{s},t_{i}, \tau_{s1}, \tau_{i1}, \tau_{s2},
\tau_{i2})\right\rangle\right\rangle_{t_s,t_i} \notag  \\
& = \left\langle{\Gamma}_{p}\left(\tau_1-\frac{t_s+t_i}{2},
\tau_2-\frac{t_s+t_i}{2}\right)\right\rangle_{\frac{t_s+t_i}{2}}\notag\\
&\quad\times\left\langle{\Gamma}_{d}\left(\tau'_1-\frac{t_s-t_i}{2},
\tau'_2-\frac{t_s-t_i}{2}\right)\right\rangle_{\frac{t_s-t_i}{2}},
 \end{align}
and then integrating it with respect to $(t_s+t_i)/2$ over $T_{\rm
pc}$ and with respect to $(t_s-t_i)/2$ over $T_{\rm ci}$. In most
experiments, the coincidence time-window $T_{\rm ci}$ spans a few
nanoseconds, which is much longer than the inverse
frequency-bandwidth of $g(\omega)$, typically of the order of
picoseconds. The photon collection time $T_{\rm pc}$ is usually a
few seconds and is much longer than the inverse
frequency-bandwidth of the pump field $V(\omega_p)$, typically of
the order of microseconds. Thus we perform the above
time-averaging in the limit $T_{pc},T_{ci}\to \infty$ to obtain
\begin{align}
{\bar\Gamma}^{(2)}& = {\bar\Gamma}_{p}\left(\tau_1,
\tau_2\right)
{\bar\Gamma}_{d}\left(\tau'_1, \tau'_2\right)\notag\\
&= \sqrt{\bar{I}_1 \bar{I}_2}\sqrt{\bar{G}_1 {\bar G}_2} \ \
{\bar\gamma}_{p}\left(\Delta\tau\right){\bar\gamma}_{d}\left(\Delta\tau'\right)\notag\\
&= \bar{R}^{(2)}{\bar\gamma}_{p}
\left(\Delta\tau\right){\bar\gamma}_{d}\left(\Delta\tau'\right).
 \end{align}
Here ${\bar I}_1={\bar\Gamma}_{p}\left(\tau_1, \tau_1\right)$,
${\bar G}_1={\bar\Gamma}_{d}\left(\tau'_1, \tau'_1\right)$,
$\bar{R}^{(2)}\equiv \sqrt{\bar{I}_1 \bar{I}_2}\sqrt{\bar{G}_1
\bar{G}_2} $, ${\bar\gamma}_{p}
\left(\Delta\tau\right)={\bar\Gamma}_{p}\left(\tau_1,
\tau_2\right)/\sqrt{\bar{I}_1 \bar{I}_2}$, and
${\bar\gamma}_{d}
\left(\Delta\tau\right)={\bar\Gamma}_{d}\left(\tau'_1,
\tau'_2\right)/\sqrt{\bar{G}_1 \bar{G}_2}$, etc. The function
$\bar{{\gamma}}_{p}(\Delta\tau)$ satisfies
$0\leq|\bar{{\gamma}}_{p}(\Delta\tau)|\leq 1$ and diminishes
over a $\Delta\tau$-scale given by the inverse pump bandwidth
$1/\Delta\omega_{p0}$. The function
$\bar{{\gamma}}_{d}(\Delta\tau')$ also satisfies
$0\leq|\bar{{\gamma}}_{d}(\Delta\tau')|\leq 1$ and diminishes
over a $\Delta\tau'$-scale given by the inverse
frequency-bandwidth $1/\Delta\omega_{d0}$. The temporal widths of
$\bar\gamma_{p}(\Delta\tau)$ and $\bar\gamma_{d}(\Delta\tau')$
limit the ranges over which fringes could be observed as functions
of $\Delta\tau'$ and $\Delta\tau$, respectively, in a
time-averaged two-photon interference experiment.

The coincidence count rate of Eq.~(\ref{interference equation}) in
the time-averaged scheme therefore becomes
\begin{multline}\label{rsi-integ}
 \bar{R}_{si}^{(2)}=\kappa_1^2 \bar{R}^{(2)}+\kappa_2^2 \bar{R}^{(2)}+
\kappa_1\kappa_2 \bar{R}^{(2)}{\bar\gamma}_{p}
\left(\Delta\tau\right){\bar\gamma}_{d}\left(\Delta\tau'\right)\\
 \times e^{i(\omega_{p0}\Delta\tau+\omega_{d0}\Delta\tau'+\Delta\phi)}+{\rm c.c.}
\end{multline}
A similar expression was reported in Ref.~\cite{jha2008pra}, where
various temporal two-photon interference effects have been
described. The time averaged coherence function
${\bar\gamma}_{p} \left(\Delta\tau\right)$ has the same
functional form as the time-averaged coherence function of the
pump field. The time-averaged coherence function
${\bar\gamma}_{d}\left(\Delta\tau'\right)$ depends on the
phase-matching function and the crystal parameters, and its
functional form shows up in the Hong-Ou-Mandel (HOM)
\cite{hong1987prl} and HOM-like effects \cite{pittman1996prl,
strekalov1998pra}.

\section{The special case of a Gaussian Schell-model pump field}

In the last section, we considered PDC with a very general
non-stationary pump field and described how the temporal coherence
properties of the pump field get transferred to the down-converted
two-photon field. In this section, we consider the pump field to
be a widely-studied class of non-stationary fields, namely, the
Gaussian Schell-model field, also known as the non-stationary
Gaussian pulsed fields \cite{paakkonen2002optcomm}.

The cross-spectral density function of a Gaussian Schell-model
field is given by \cite{paakkonen2002optcomm}
\begin{multline}\label{GS-cross correlation}
\hspace{-4mm}\big\langle V^{*}(\omega''+
\omega_{0})V(\omega'+\omega_{0})\big\rangle=\\
A\exp\left[-\frac{({\omega'}^{2}+
{\omega''}^{2})}{4\left(\Delta\omega_{p0}\right)^{2}}\right]
\exp\left[-\frac{\left(\omega'-\omega''\right)^{2}}{2
\left(\Delta\omega_{c}\right)^{2}}\right].,
\end{multline}
where $\Delta\omega_{p0}$ is the frequency bandwidth of the field.
The parameter $\Delta\omega_{c}$ is called the spectral
correlation width and it quantifies the frequency-separation up to
which different frequency components are phase-correlated. The
limit $\Delta\omega_{c}\to0$ corresponds to a continuous-wave,
stationary field in which case the constituent frequency
components are completely uncorrelated. The other limit
$\Delta\omega_{c}\to\infty$ corresponds to a fully-coherent pulsed
field in which case the constituent frequency components are
perfectly phase-correlated. The corresponding temporal correlation
function ${\Gamma}_{\rm GS}(t_1, t_2)$ can be calculated by
using the generalized Wiener-Khintchine theorem
\cite{paakkonen2002optcomm}:
\begin{align}
{\Gamma}_{\rm GS}(t_1, t_2)=\sqrt{I(t_1)I(t_2)}\gamma_{\rm
GS}(\Delta t),\label{gs-model pump}
\end{align}
with $\Delta t=t_1-t_2$ and where
\begin{align}
I(t_{1(2)})&=\frac{2\pi \Delta\omega_{p0}
A}{T}\exp\left[-\frac{t_{1(2)}^2}{2T^2}\right],  \notag\\
{\rm and}\qquad{\gamma}_{\rm GS}(\Delta
t)&=\exp\left[-\frac{(\Delta t)^{2}}{2\tau_{\rm
coh}^{2}}\right].\notag
\end{align}
Here $\tau_{\rm
coh}=(\Delta\omega_{c}/\Delta\omega_{p0})\left[1/(2\Delta
\omega_{p0})^{2}+1/(\Delta\omega_{c})^{2}\right]^{1/2}$ is a
measure of the coherence time of the field and $T=\left[1/(2\Delta
\omega_{p0})^{2}+1/(\Delta\omega_{c})^{2}\right]^{1/2}$ is a
measure of the temporal width of the non-stationary Gaussian
pulse. The limit $\Delta\omega_{c}\to\infty$ yields $\tau_{\rm
coh}\to\infty$ as expected for a fully-coherent field, and the
other limit $\Delta\omega_{c}\to0$ yields $\tau_{\rm
coh}=1/\Delta\omega_{p0}$ as expected for a continuous-wave,
stationary field.

Now, for conceptual clarity, we assume in this section that
${\Gamma}_{d}\left(\tau'_1-\frac{t_s-t_i}{2},
\tau'_2-\frac{t_s-t_i}{2}\right)=1$ and take the pump field to be
the Gaussian Schell-model field given by Eq.~(\ref{GS-cross
correlation}). Eq.~(\ref{r12-decomp}) then becomes:
\begin{align}
&{\Gamma}^{(2)}(t_s, t_i,\tau_1,
\tau_2)={\Gamma}_{p}\left(\tau_1-\frac{t_s+t_i}{2},
\tau_2-\frac{t_s+t_i}{2}\right)\notag\\
&=\sqrt{I\left(\tau_1-\frac{t_s+t_i}{2}\right)I\left(\tau_2-\frac{t_s+t_i}{2}\right)}\gamma_p(\Delta\tau),
\label{gs-model two-photon}
\end{align}
where
\begin{align}
&I\left(\tau_{1}-\frac{t_s+t_i}{2}\right)=\frac{2\pi
\Delta\omega_{p0}
A}{T}\exp\left[-\frac{\left(\tau_{1}-\frac{t_s+t_i}{2}\right)^2}{2T^2}\right], \ {\rm etc,}\notag\\
&{\rm and} \quad
{\gamma}_{p}(\Delta\tau)=\exp\left[-\frac{\Delta\tau^{2}}{2\tau_{\rm
coh}^{2}}\right].\notag
\end{align}
As expected from Eq.~(\ref{r12-decomp}), we find that in terms of
$\tau_1-\frac{t_s+t_i}{2}$ and $\tau_2-\frac{t_s+t_i}{2}$, the
two-photon cross-correlation function in Eq.~(\ref{gs-model
two-photon}) assumes the same functional form as does the
cross-correlation function in Eq.~(\ref{gs-model pump}) in terms
of  $t_1$ and $t_2$. When integrated over $t$, Eq.~(\ref{gs-model
two-photon}) yields
\begin{align}
{\bar\Gamma}^{(2)}={\bar\Gamma}_{p}(\tau_1,
\tau_2)&=\sqrt{\bar{I}_1 \bar{I}_2}\bar\gamma_p(\Delta\tau),
\end{align}
with $\bar{I}_1 =\bar{I}_2=
(2\pi)^{\frac{3}{2}}\Delta\omega_{p0}A$ and
\begin{align}
{\bar\gamma}_{p}(\Delta\tau)=\exp\left[-\frac{\Delta\tau^{2}}{2\bar\tau_{\rm
coh}^{2}}\right],\notag
\end{align}
where $\bar{\tau}_{\rm coh}=1/\Delta\omega_{p0}$ is a measure of
the coherence time. The time averaging washes out effects due to
frequency correlations. Thus, only in the case of a stationary
pump field $\bar\gamma_p(\Delta\tau)=\gamma_p(\Delta\tau)$ and
$\bar{\tau}_{\rm coh}=\tau_{\rm coh}$.

\

\section{Pump temporal coherence and two-qubit ENERGY-TIME ENTANGLEMENT}

Two-qubit states are the necessary ingredients for many quantum
information protocols \cite{ekert1991prl, bennett1992prl,
bennett1993prl} and have been realized by exploiting the
entanglement of PDC photons in several degrees of freedom
including polarization \cite{kwiat1995prl}, time-energy
\cite{franson1989prl, brendel1999prl, tittel2000prl, thew2002pra,
ramelow2009arxiv, kwon2009optexp, kwon2013optexp},
position-momentum \cite{neves2005prl, neves2007pra,
rarity1990prl}, and orbital angular momentum (OAM)
\cite{vaziri2002prl, langford2004prl, leach2009optexp,
jha2010prl}. There have been previous studies describing how
correlations of the pump field in polarization and spatial degrees
of freedom affect the entanglement of the generated two-qubit
states. In the polarization degree of freedom it was shown
\cite{kulkarni2016pra} that the degree of polarization $P$ of the
pump photon puts an upper bound of $(1+P)/2$ on the concurrence of
the generated two-qubit state. In the spatial degree of freedom,
effects of pump spatial coherence on the entanglement of the
generated spatial two-qubit state have been worked out for
two-qubit state that have only two non-zero diagonal elements, and
for such states it has been shown that the concurrence is bounded
by the degree of spatial coherence of the pump field
\cite{jha2010pra}. However, to the best of our knowledge, no such
relation has so far been derived for the time-energy entangled
two-qubit states.

There are two generic methods by which one makes a PDC-based
time-energy entangled two-qubit state. In the first method, one
uses a continuous-wave pump field, either single-mode
\cite{franson1989prl} or multi-mode \cite{kwon2009optexp,
kwon2013optexp}. In the second method, one uses a pulsed pump
field \cite{brendel1999prl, tittel2000prl, thew2002pra,
ramelow2009arxiv}. In both these methods, a combination of
post-selection strategies, such as selecting a faster coincidence
detection-window, using arrival time of pump photon as a trigger
etc., one makes sure that there are only two alternative pathways
in which the signal and idler photons reach their respective
detectors. The two alternative pathways form the two dimensional
qubit space for the signal and idler photons. We represent by
$|s1\rangle$ the state of the signal photon in alternative 1, etc.
Therefore, the density matrix $\rho_{\rm 2qubit}$ of the two-qubit
state can be written in the basis \{$|s1\rangle|i1\rangle,
|s1\rangle|i2\rangle, |s2\rangle|i1\rangle,
|s2\rangle|i2\rangle$\} as:
\begin{equation}\label{2qubitstate}
\rho=\begin{pmatrix}
a  &&  0  &&  0  &&  c \\
0  &&  0  &&  0  &&  0 \\
0  &&  0  &&  0  &&  0 \\
c^*  &&  0  &&  0  && b
\end{pmatrix}
\end{equation}
where the diagonal terms $a$ and $b$ are the probabilities that
the signal and idler photons are detected in states
$|s1\rangle|i1\rangle$ and $|s2\rangle|i2\rangle$, respectively,
and the off-diagonal term $c$ is a measure of coherence between
states $|s1\rangle|i1\rangle$ and $|s2\rangle|i2\rangle$. In an
experimental situation, the density matrix $\rho$ can be
represented by the two alternative pathways of
Fig.~\ref{two-photon-path-diagram}. Therefore, using
Eq.~(\ref{rsi-integ}), we write $ a=\eta\kappa_1^2 \bar{R}^{(2)}$
and $b=\eta\kappa_2^2 \bar{R}^{(2)}$, where $\eta=1/[\kappa_1^2
\bar{R}^{(2)}+\kappa_2^2 \bar{R}^{(2)}]$ is the constant of
proportionality. The off-diagonal term is given by
\begin{align}
\!c=\eta\kappa_1\kappa_2 \bar{R}^{(2)}{\bar\gamma}_{p}
(\Delta\tau){\bar\gamma}_{d}(\Delta\tau')
 e^{i(\omega_{p0}\Delta\tau+\omega_{d0}\Delta\tau'+\Delta\phi)}.
\end{align}
The entanglement of $\rho_{\rm 2qubit}$, as quantified by
Wootters's concurrence $C(\rho_{\rm 2qubit})$
\cite{wootters1998prl}, can be shown to be
\begin{align}\label{conc1}
\!\!\!\!\!C(\rho_{\rm 2qubit})=2|c|=\frac{2\kappa_1\kappa_2
\bar{R}^{(2)}}{\kappa_1^2 \bar{R}^{(2)}+\kappa_2^2
\bar{R}^{(2)}}{\bar\gamma}_{p}
(\Delta\tau){\bar\gamma}_{d}(\Delta\tau').
\end{align}
The pre-factor $2 \kappa_1\kappa_2 \bar{R}^{(2)}/(\kappa_1^2
\bar{R}^{(2)}+\kappa_2^2 \bar{R}^{(2)})$ is no greater than 1, and
${\bar\gamma}_{d}(\Delta\tau')$ also satisfies
$0\leq|{\bar\gamma}_{d}(\Delta\tau')|\leq 1$. We therefore
arrive at the relation: $
 C(\rho_{\rm 2qubit})\leq \bar{{\gamma}}_{p}(\Delta\tau)$.
 Therefore, we find that the concurrence $C(\rho_{\rm 2qubit})$ of
the time-energy two-qubit state is bounded from above by the
degree of coherence of the pump photon and thus that the temporal
correlations of the pump field set an upper bound on the
attainable concurrence for a two-qubit state of the form of
Eq.~(\ref{2qubitstate}). We note that in situations in which ${\bar\gamma}_{d}(\Delta\tau')\approx 1$ and $\kappa_1=\kappa_2$, the maximum achievable concurrence for a pulsed field can be unity in principle and for a continuous-wave field it can be unity as long as $\Delta\tau$ is much smaller than the coherence time of the pump field. The above result is the temporal analog of the
results obtained in the polarization \cite{kulkarni2016pra} and
spatial \cite{jha2010pra} degrees of freedom. However, unlike in
the spatial degree of freedom, which does not involve any
space-averaged detection scheme, the results derived in this
article show that even for the time-averaged detection schemes,
the temporal correlation properties of the pump do directly decide
the upper limit on entanglement that a time-energy entangled
two-qubit state can achieve.

\section{CONCLUSIONS AND DISCUSSIONS}

In conclusions, we have shown that in parametric down-conversion
the coherence properties of a temporally partially coherent pump
field get entirely transferred to the down-converted entangled
two-photon field. Under the assumption that the
frequency-bandwidth of the down-converted signal-idler photons is
much larger than that of the pump, we have worked out the temporal
coherence functions of the down-converted field for both
infinitely-fast and time-averaged detection schemes. We have shown
that in each scheme the coherence function factorizes into two
separate coherence functions with one of them carrying the entire
statistical information of the pump field. Taking the pump to be a
Gaussian Schell-model field, we have derived explicit expressions
for the coherence functions. Finally, we have shown that the
concurrence of time-energy entangled two-qubit states is bounded
by the degree of temporal coherence of the pump field. This result
extends previously obtained results in the spatial
\cite{jha2010pra} and polarization \cite{kulkarni2016pra} degrees
of freedom to the temporal degree of freedom and can thus have
important implications for understanding how correlations of the
pump field in general manifest as two-particle entanglement. Our
results can also be important for time-energy two-qubit based
quantum communication applications. This is because it has been
recognized that energy-time entangled two-qubit states are better
than the polarization two-qubit states for long-distance quantum
information \cite{marcikic2003nature, marcikic2004prl}, and our
results show that the temporal coherence properties of the pump
field can be used as a parameter for tailoring the two-qubit
time-energy entanglement. Moreover, it is known that the purity of the individual photon states increases with the decrease in the entanglement of the two-photon state. Therefore, our work can also have implications for
PDC-based heralded single photons sources \cite{obrien2007science, cassemiro2010njp} in the sense that the degree of
purity of heralded photons can be tailored by controlling the
coherence properties of the pump field.

\section*{ACKNOWLEDGMENTS}

We gratefully acknowledge financial support through an initiation
grant no. IITK /PHY /20130008 from Indian Institute of Technology
(IIT) Kanpur, India and through the research grant no.
EMR/2015/001931 from the Science and Engineering Research Board
(SERB), Department of Science and Technology, Government of India.


\begin{thebibliography}{10}
\newcommand{\enquote}[1]{``#1''}

\bibitem{vogel2014pra}
W.~Vogel and J.~Sperling, \enquote{Unified quantification of nonclassicality
  and entanglement,} Phys. Rev. A \textbf{89}, 052302 (2014).

\bibitem{baumgratz2014prl}
T.~Baumgratz, M.~Cramer, and M.~B. Plenio, \enquote{Quantifying coherence,}
  Phys. Rev. Lett. \textbf{113}, 140401 (2014).

\bibitem{girolami2014prl}
D.~Girolami, \enquote{Observable measure of quantum coherence in finite
  dimensional systems,} Phys. Rev. Lett. \textbf{113}, 170401 (2014).

\bibitem{chitambar2016prl}
E.~Chitambar and M.-H. Hsieh, \enquote{Relating the resource theories of
  entanglement and quantum coherence,} Phys. Rev. Lett. \textbf{117}, 020402
  (2016).

\bibitem{burnham1970prl}
D.~C. Burnham and D.~L. Weinberg, \enquote{Observation of simultaneity in
  parametric production of optical photon pairs,} Phys. Rev. Lett. \textbf{25},
  84--87 (1970).

\bibitem{brendel1995pra}
J.~Brendel, W.~Dultz, and W.~Martienssen, \enquote{Geometric phases in
  two-photon interference experiments,} Phys. Rev. A \textbf{52}, 2551--2556
  (1995).

\bibitem{hong1987prl}
C.~K. Hong, Z.~Y. Ou, and L.~Mandel, \enquote{Measurement of subpicosecond time
  intervals between two photons by interference,} Phys. Rev. Lett. \textbf{59},
  2044--2046 (1987).

\bibitem{zou1991prl}
X.~Y. Zou, L.~J. Wang, and L.~Mandel, \enquote{Induced coherence and
  indistinguishability in optical interference,} Phys. Rev. Lett. \textbf{67},
  318--321 (1991).

\bibitem{herzog1994prl}
T.~J. Herzog, J.~G. Rarity, H.~Weinfurter, and A.~Zeilinger,
  \enquote{Frustrated two-photon creation via interference,} Phys. Rev. Lett.
  \textbf{72}, 629--632 (1994).

\bibitem{pittman1996prl}
T.~B. Pittman, D.~V. Strekalov, A.~Migdall, M.~H. Rubin, A.~V. Sergienko, and
  Y.~H. Shih, \enquote{Can two-photon interference be considered the
  interference of two photons?} Phys. Rev. Lett. \textbf{77}, 1917--1920
  (1996).

\bibitem{jha2008pra}
A.~K. Jha, M.~N. O'Sullivan, K.~W.~C. Chan, and R.~W. Boyd, \enquote{Temporal
  coherence and indistinguishability in two-photon interference effects,} Phys.
  Rev. A \textbf{77}, 021801(R) (2008).

\bibitem{franson1989prl}
J.~D. Franson, \enquote{Bell inequality for position and time,} Phys. Rev.
  Lett. \textbf{62}, 2205--2208 (1989).

\bibitem{brendel1991prl}
J.~Brendel, E.~Mohler, and W.~Martienssen, \enquote{Time-resolved dual-beam
  two-photon interferences with high visibility,} Phys. Rev. Lett. \textbf{66},
  1142--1145 (1991).

\bibitem{jha2008prl}
A.~K. Jha, M.~Malik, and R.~W. Boyd, \enquote{Exploring energy-time
  entanglement using geometric phase,} Phys. Rev. Lett. \textbf{101}, 180405
  (2008).

\bibitem{brendel1999prl}
J.~Brendel, N.~Gisin, W.~Tittel, and H.~Zbinden, \enquote{Pulsed energy-time
  entangled twin-photon source for quantum communication,} Phys. Rev. Lett.
  \textbf{82}, 2594--2597 (1999).

\bibitem{thew2002pra}
R.~T. Thew, S.~Tanzilli, W.~Tittel, H.~Zbinden, and N.~Gisin,
  \enquote{Experimental investigation of the robustness of partially entangled
  qubits over 11 km,} Phys. Rev. A \textbf{66}, 062304 (2002).

\bibitem{fonseca1999pra}
E.~J.~S. Fonseca, C.~H. Monken, S.~P\'adua, and G.~A. Barbosa,
  \enquote{Transverse coherence length of down-converted light in the
  two-photon state,} Phys. Rev. A \textbf{59}, 1608--1614 (1999).

\bibitem{neves2007pra}
L.~Neves, G.~Lima, E.~J.~S. Fonseca, L.~Davidovich, and S.~P\'{a}dua,
  \enquote{Characterizing entanglement in qubits created with spatially
  correlated twin photons,} Phys. Rev. A \textbf{76}, 032314 (2007).

\bibitem{jha2010pra}
A.~K. Jha and R.~W. Boyd, \enquote{Spatial two-photon coherence of the
  entangled field produced by down-conversion using a partially spatially
  coherent pump beam,} Phys. Rev. A \textbf{81}, 013828 (2010).

\bibitem{nagali2009natphot}
E.~Nagali, L.~Sansoni, F.~Sciarrino, F.~De~Martini, L.~Marrucci, B.~Piccirillo,
  E.~Karimi, and E.~Santamato, \enquote{Optimal quantum cloning of orbital
  angular momentum photon qubits through hong--ou--mandel coalescence,} Nature
  Photonics \textbf{3}, 720--723 (2009).

\bibitem{jha2010prl}
A.~K. Jha, J.~Leach, B.~Jack, S.~Franke-Arnold, S.~M. Barnett, R.~W. Boyd, and
  M.~J. Padgett, \enquote{Angular two-photon interference and angular two-qubit
  states,} Phys. Rev. Lett. \textbf{104}, 010501 (2010).

\bibitem{pires2010prl}
H.~Di~Lorenzo~Pires, H.~C.~B. Florijn, and M.~P. van Exter,
  \enquote{Measurement of the spiral spectrum of entangled two-photon states,}
  Phys. Rev. Lett. \textbf{104}, 020505 (2010).

\bibitem{jha2011pra}
A.~K. Jha, G.~S. Agarwal, and R.~W. Boyd, \enquote{Partial angular coherence
  and the angular schmidt spectrum of entangled two-photon fields,} Phys. Rev.
  A \textbf{84}, 063847 (2011).

\bibitem{hong1985pra}
C.~K. Hong and L.~Mandel, \enquote{Theory of parametric frequency down
  conversion of light,} Phys. Rev. A \textbf{31}, 2409--2418 (1985).

\bibitem{rubin1996pra}
M.~H. Rubin, \enquote{Transverse correlation in optical spontaneous parametric
  down-conversion,} Phys. Rev. A \textbf{54}, 5349--5360 (1996).

\bibitem{monken1998pra}
C.~H. Monken, P.~H.~S. Ribeiro, and S.~P\'adua, \enquote{Transfer of angular
  spectrum and image formation in spontaneous parametric down-conversion,}
  Phys. Rev. A \textbf{57}, 3123--3126 (1998).

\bibitem{joobeur1996pra}
A.~Joobeur, B.~E.~A. Saleh, T.~S. Larchuk, and M.~C. Teich, \enquote{Coherence
  properties of entangled light beams generated by parametric down-conversion:
  Theory and experiment,} Phys. Rev. A \textbf{53}, 4360--4371 (1996).

\bibitem{ribeiro1997pra}
P.~H. Souto~Ribeiro, \enquote{Partial coherence with twin photons,} Phys. Rev.
  A \textbf{56}, 4111--4117 (1997).

\bibitem{saleh2005prl}
B.~E.~A. Saleh, M.~C. Teich, and A.~V. Sergienko, \enquote{Wolf equations for
  two-photon light,} Phys. Rev. Lett. \textbf{94}, 223601 (2005).

\bibitem{walborn2004pra}
S.~P. Walborn, A.~N. de~Oliveira, R.~S. Thebaldi, and C.~H. Monken,
  \enquote{Entanglement and conservation of orbital angular momentum in
  spontaneous parametric down-conversion,} Phys. Rev. A \textbf{69}, 023811
  (2004).

\bibitem{kulkarni2016pra}
G.~Kulkarni, V.~Subrahmanyam, and A.~K. Jha, \enquote{Intrinsic upper bound on
  two-qubit polarization entanglement predetermined by pump polarization
  correlations in parametric down-conversion,} Physical Review A \textbf{93},
  063842 (2016).

\bibitem{grice1997pra}
W.~P. Grice and I.~A. Walmsley, \enquote{Spectral information and
  distinguishability in type-{II} down-conversion with a broadband pump,} Phys.
  Rev. A \textbf{56}, 1627--1634 (1997).

\bibitem{keller1997pra}
T.~E. Keller and M.~H. Rubin, \enquote{Theory of two-photon entanglement for
  spontaneous parametric down-conversion driven by a narrow pump pulse,} Phys.
  Rev. A \textbf{56}, 1534--1541 (1997).

\bibitem{tittel2000prl}
W.~Tittel, J.~Brendel, H.~Zbinden, and N.~Gisin, \enquote{Quantum cryptography
  using entangled photons in energy-time bell states,} Phys. Rev. Lett.
  \textbf{84}, 4737--4740 (2000).

\bibitem{mikhailova2008pra}
Y.~M. Mikhailova, P.~A. Volkov, and M.~V. Fedorov, \enquote{Biphoton wave
  packets in parametric down-conversion: Spectral and temporal structure and
  degree of entanglement,} Phys. Rev. A \textbf{78}, 062327 (2008).

\bibitem{inagaki2013optexp}
T.~Inagaki, N.~Matsuda, O.~Tadanaga, M.~Asobe, and H.~Takesue,
  \enquote{Entanglement distribution over 300 km of fiber,} Opt. Express
  \textbf{21}, 23241--23249 (2013).

\bibitem{ou1989pra}
Z.~Y. Ou, L.~J. Wang, and L.~Mandel, \enquote{Vacuum effects on interference in
  two-photon down conversion,} Phys. Rev. A \textbf{40}, 1428--1435 (1989).

\bibitem{rubin1994pra}
M.~H. Rubin, D.~N. Klyshko, Y.~H. Shih, and A.~V. Sergienko, \enquote{Theory of
  two-photon entanglement in type-{II} optical parametric down-conversion,}
  Phys. Rev. A \textbf{50}, 5122--5133 (1994).

\bibitem{milonni1996pra}
P.~W. Milonni, H.~Fearn, and A.~Zeilinger, \enquote{Theory of two-photon
  down-conversion in the presence of mirrors,} Phys. Rev. A \textbf{53},
  4556--4566 (1996).

\bibitem{kwon2009optexp}
O.~Kwon, Y.-S. Ra, and Y.-H. Kim, \enquote{Coherence properties of spontaneous
  parametric down-conversion pumped by a multi-mode cw diode laser,} Optics
  express \textbf{17}, 13059--13069 (2009).

\bibitem{nasr2008prl}
M.~B. Nasr, S.~Carrasco, B.~E.~A. Saleh, A.~V. Sergienko, M.~C. Teich, J.~P.
  Torres, L.~Torner, D.~S. Hum, and M.~M. Fejer, \enquote{Ultrabroadband
  biphotons generated via chirped quasi-phase-matched optical parametric
  down-conversion,} Phys. Rev. Lett. \textbf{100}, 183601 (2008).

\bibitem{okano2015scirep}
M.~Okano, H.~H. Lim, R.~Okamoto, N.~Nishizawa, S.~Kurimura, and S.~Takeuchi,
  \enquote{0.54 $\mu$m resolution two-photon interference with dispersion
  cancellation for quantum optical coherence tomography,} Scientific reports
  \textbf{5} (2015).

\bibitem{tanaka2012optexp}
A.~Tanaka, R.~Okamoto, H.~H. Lim, S.~Subashchandran, M.~Okano, L.~Zhang,
  L.~Kang, J.~Chen, P.~Wu, T.~Hirohata, S.~Kurimura, and S.~Takeuchi, \enquote{Noncollinear
  parametric fluorescence by chirped quasi-phase matching for monocycle
  temporal entanglement,} Optics express \textbf{20}, 25228--25238 (2012).

\bibitem{paakkonen2002optcomm}
P.~P\"{a}\"{a}kk\"{o}nen, J.~Turunen, P.~Vahimaa, A.~T. Friberg, and
  F.~Wyrowski, \enquote{Partially coherent {G}aussian pulses,} Opt. Comm.
  \textbf{204}, 53 -- 58 (2002).

\bibitem{wang1991pra}
L.~J. Wang, X.~Y. Zou, and L.~Mandel, \enquote{Induced coherence without
  induced emission,} Phys. Rev. A \textbf{44}, 4614--4622 (1991).

\bibitem{glauber1963pr}
R.~J. Glauber, \enquote{The quantum theory of optical coherence,} Phys. Rev.
  \textbf{130}, 2529--2539 (1963).

\bibitem{kwon2013optexp}
O.~Kwon, K.-K. Park, Y.-S. Ra, Y.-S. Kim, and Y.-H. Kim, \enquote{Time-bin
  entangled photon pairs from spontaneous parametric down-conversion pumped by
  a cw multi-mode diode laser,} Optics express \textbf{21}, 25492--25500
  (2013).

\bibitem{marcikic2002pra}
I.~Marcikic, H.~de~Riedmatten, W.~Tittel, V.~Scarani, H.~Zbinden, and N.~Gisin,
  \enquote{Time-bin entangled qubits for quantum communication created by
  femtosecond pulses,} Phys. Rev. A \textbf{66}, 062308 (2002).

\bibitem{marcikic2003nature}
I.~Marcikic, H.~De~Riedmatten, W.~Tittel, H.~Zbinden, and N.~Gisin,
  \enquote{Long-distance teleportation of qubits at telecommunication
  wavelengths,} Nature \textbf{421}, 509--513 (2003).

\bibitem{marcikic2004prl}
I.~Marcikic, H.~de~Riedmatten, W.~Tittel, H.~Zbinden, M.~Legr\'e, and N.~Gisin,
  \enquote{Distribution of time-bin entangled qubits over 50 km of optical
  fiber,} Phys. Rev. Lett. \textbf{93}, 180502 (2004).

\bibitem{mandel&wolf}
L.~Mandel and E.~Wolf, \emph{Optical Coherence and Quantum Optics} (Cambridge
  university press, New York, 1995).

\bibitem{jha2010pra2}
A.~K. Jha, G.~A. Tyler, and R.~W. Boyd, \enquote{Effects of atmospheric
  turbulence on the entanglement of spatial two-qubit states,} Phys. Rev. A
  \textbf{81}, 053832 (2010).

\bibitem{strekalov1998pra}
D.~V. Strekalov, T.~B. Pittman, and Y.~H. Shih, \enquote{What we can learn
  about single photons in a two-photon interference experiment,} Phys. Rev. A
  \textbf{57}, 567--570 (1998).

\bibitem{ekert1991prl}
A.~K. Ekert, \enquote{Quantum cryptography based on {B}ell\char39{}s theorem,}
  Phys. Rev. Lett. \textbf{67}, 661--663 (1991).

\bibitem{bennett1992prl}
C.~H. Bennett and S.~J. Wiesner, \enquote{Communication via one- and
  two-particle operators on {E}instein-{P}odolsky-{R}osen states,} Phys. Rev.
  Lett. \textbf{69}, 2881--2884 (1992).

\bibitem{bennett1993prl}
C.~H. Bennett, G.~Brassard, C.~Cr\'epeau, R.~Jozsa, A.~Peres, and W.~K.
  Wootters, \enquote{Teleporting an unknown quantum state via dual classical
  and {E}instein-{P}odolsky-{R}osen channels,} Phys. Rev. Lett. \textbf{70},
  1895--1899 (1993).

\bibitem{kwiat1995prl}
P.~G. Kwiat, K.~Mattle, H.~Weinfurter, A.~Zeilinger, A.~V. Sergienko, and
  Y.~Shih, \enquote{New high-intensity source of polarization-entangled photon
  pairs,} Phys. Rev. Lett. \textbf{75}, 4337--4341 (1995).

\bibitem{ramelow2009arxiv}
S.~Ramelow, L.~Ratschbacher, A.~Fedrizzi, N.~Langford, and A.~Zeilinger,
  \enquote{Discrete, tunable color entanglement,} Phys. Rev. Lett. \textbf{103}, 253601 (2009).

\bibitem{neves2005prl}
L.~Neves, G.~Lima, J.~G. Aguirre~G\'omez, C.~H. Monken, C.~Saavedra, and
  S.~P\'adua, \enquote{Generation of entangled states of qudits using twin
  photons,} Phys. Rev. Lett. \textbf{94}, 100501 (2005).

\bibitem{rarity1990prl}
J.~G. Rarity and P.~R. Tapster, \enquote{Experimental violation of
  {B}ell\char39{}s inequality based on phase and momentum,} Phys. Rev. Lett.
  \textbf{64}, 2495--2498 (1990).

\bibitem{vaziri2002prl}
A.~Vaziri, G.~Weihs, and A.~Zeilinger, \enquote{Experimental two-photon,
  three-dimensional entanglement for quantum communication,} Phys. Rev. Lett.
  \textbf{89}, 240401 (2002).

\bibitem{langford2004prl}
N.~K. Langford, R.~B. Dalton, M.~D. Harvey, J.~L. O\char39{}Brien, G.~J. Pryde,
  A.~Gilchrist, S.~D. Bartlett, and A.~G. White, \enquote{Measuring entangled
  qutrits and their use for quantum bit commitment,} Phys. Rev. Lett.
  \textbf{93}, 053601 (2004).

\bibitem{leach2009optexp}
J.~Leach, B.~Jack, J.~Romero, M.~Ritsch-Marte, R.~W. Boyd, A.~K. Jha, S.~M.
  Barnett, S.~Franke-Arnold, and M.~J. Padgett, \enquote{Violation of a {B}ell
  inequality in two-dimensional orbital angular momentum state-spaces,} Opt.
  Express \textbf{17}, 8287--8293 (2009).

\bibitem{wootters1998prl}
W.~K. Wootters, \enquote{Entanglement of formation of an arbitrary state of two
  qubits,} Phys. Rev. Lett. \textbf{80}, 2245--2248 (1998).

\bibitem{obrien2007science}
J.~L. O'Brien, \enquote{Optical quantum computing,} Science \textbf{318},
  1567--1570 (2007).

\bibitem{cassemiro2010njp}
K.~N. Cassemiro, K.~Laiho, and C.~Silberhorn, \enquote{Accessing the purity of
  a single photon by the width of the hong--ou--mandel interference,} New
  Journal of Physics \textbf{12}, 113052 (2010).

\end{thebibliography}

\end{document}